\documentclass[12pt,twoside]{article}
\usepackage[T1]{fontenc}
\usepackage[latin1]{inputenc}
\usepackage{times}
\usepackage{graphicx}
\usepackage{a4wide}
\usepackage{listings}
 \usepackage{amsmath}

\begin{document}

\begin{center}

\section*{The TRB for HADES and FAIR experiments at GSI\footnote{
    Proceedings of the 10th Conference on Astroparticle, Particle and Space Physics, Detectors and Medical Physics Applications, 973-977, 2008 
  }}

\thispagestyle{empty}

\markboth{W.~KRZEMIEN {\it et al.}}
{The TRB for HADES and FAIR experiments at GSI}

{I. FR\"OHLICH, J. MICHEL, C. SCHRADER, H. STR\"OBELE, J. STROTH, A.TARANTOLA} \\ \vspace{0.3cm}

{Institut f\"ur Kernphysik, Goethe-Universit\"at,\\
60486 Frankfurt, Germany\\}\vspace{0.5cm}

{M. KAJETANOWICZ}\\ \vspace{0.3cm}

{Nowoczesna Elektronika\\
  Pl-30-109 Cracow, Poland\\}\vspace{0.5cm}

{K. KORCYL, {\underline{W.~KRZEMIEN$^*$}}, M. PALKA, P. SALABURA, R. TREBACZ}\\ \vspace{0.3cm}

{Institute of Physics, Jagiellonian University,\\
Pl-30-059 Cracow, Poland\\
$^*$E-mail: wojciech.krzemien@if.uj.edu.pl} \\ \vspace{0.5cm}

{P. SKOTT, M. TRAXLER}\\ \vspace{0.3cm}

{
GSI 
Helmholtzzentrum f\"ur Schwerionenforschung GmbH\\
64291 Darmstadt, Germany\\}\vspace{0.5cm}

\textbf{Abstract}
\end{center}

The TRB hardware module is a multi-purpose Trigger and Readout Board with
on-board DAQ functionality developed for the upgrade of the HADES experiment.
It contains a single computer chip (Etrax) running Linux as a well as a 100
Mbit/s Ethernet interface.  It has been orginally designed to work as a  128-channel
Time to Digital Converter based on the HPTDC chip from CERN.  The new version
contains a 2~Gbit/s optical link and an interface connector (15 Gbit/s)
in order to realize an add-on card concept which makes the hardware very flexible.
Moreover, an FPGA chip (Xilinx, Virtex 4 LX 40) and a TigerSharc DSP provide new
computing resources which can be used to run on-line analysis algorithms.  The
TRB is proposed as a prototype for new  modules for the planned detector
systems PANDA and CBM at the future FAIR facility at GSI-Darmstadt.

{\it Keywords: Triggering; Data acquisition.}

\section{Introduction}

HADES is a high resolution dilepton spectrometer for hadron and heavy ion
physics at the SIS18 accelerator facility of GSI, Germany \cite{had}.  Its
main goal is the study of in-medium modification of light vector mesons
properties.  PANDA \cite{pan} and CBM \cite{cbm} are detector systems which
will operate at the new Facility of Antiproton and Ion Research (FAIR) at GSI.
The large research program of PANDA and CBM will address the questions of the
generation of mass, spontaneous breaking of chiral symmetry, the limits of
hadronic existence and the transition to de-confined matter \cite{jo}.  To
provide access to rare decay channels the experiments will be operated at high
luminosity. The cross sections and branching ratios of interesting processes
are low and the background will be of orders of magnitude higher in yields.
Therefore, to collect statistically relevant data, the detector systems should
operate with interaction rates above 10 MHz, which is a real challenge for the
data acquisition systems. HADES has worked out a two-level trigger
system whereas PANDA and CBM plan to design their DAQ systems without any
central trigger. Thus, data generated by the PANDA and CBM subdetectors will
be time stamped and sent to the event buildind system. In all detector systems
mentioned above, a large amount of data has to be handled which calls for high
level data reduction mechanisms.  In this context, a universal readout
electronics for various detectors  has been developed.  The ``Trigger and Readout
Board'' (TRB) is a general-purpose electronic device which can be used as base
readout module for the development of the future DAQ systems in the FAIR
experiments.

\section{The TRB version 1}

TRB version 1 (TRBv1) has been orginally developed as a 128-channel Time to
Digital Converter based on the HPTDC~\cite{tdc} chips.

\begin{figure}
\begin{center}\includegraphics[width=0.6\textwidth]{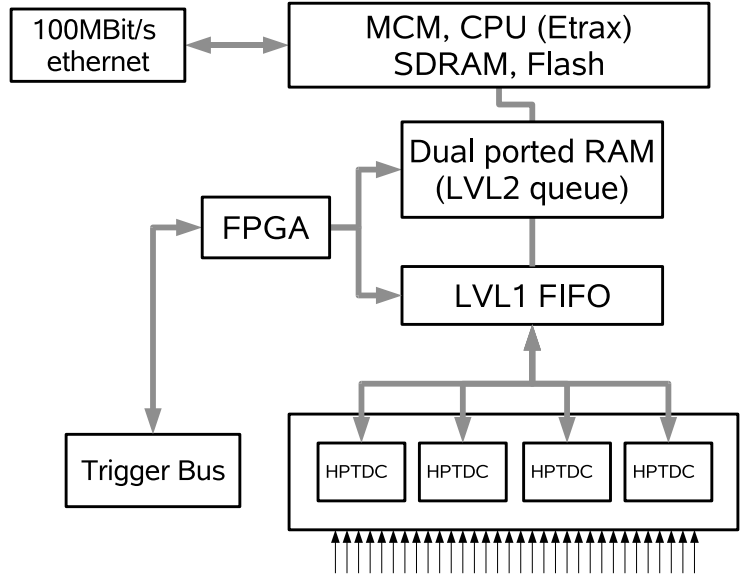}
\end{center}
\caption{TRBv1 block diagram. The readout of the data is started via a
  signal on the trigger bus. The data is stored in the FIFO queue until the
  LVL2 decision arrives. Final data formatting is done by Etrax
  chip. (Fig. taken from~\cite{tri}) }
\end{figure}

It contains 4-input connectors (high density 80 pins). Each connector permits
to sample 31 LVDS timing signals. Four 32-channel HPTDC chips perform time to
digital conversion. The 32th channel of each HPTDC is connected to the
external reference timing signal.  The HPTDC chips are highly configurable
\cite{tdc}.  The TDC binnig could be chosen in the range of 25 to 780 ps. The
HPTDC allows to measure rising and falling edges, and to define a matching
window. Moreover, the TRBv1 contains an FPGA (Xilinx Spartan 2), a FIFO queue
and dual ported RAM memory.  The slow control and the readout functionality
(for HADES) is done via an Etrax single chip computer running the Linux
operation system \cite{tre}.  The data acquisition is started by an external
level one trigger (LVL1). The FPGA controller initiates the readout and
selected data from HPTDC is stored in the FIFO queue upon the level two
trigger (LVL2) has been received.  If the LVL2 signal is positive data is
moved to the dual ported RAM.  The data is consequently read out by the Etrax and send
via UDP protocol over 100 Mbit/s Ethernet to the event builder.

The time resolution between two channels from different HPTDC chips resulted
40 ps (with 100 ps binning) \cite{tri}.  The crosstalk influence to resolution
has been estimated to be less than 20 ps.  The maximum data transport rate via
UDP has been measured to be 8 MB/s \cite{tre}.  The maximum LVL2 trigger rate was
about 18 kHz.  The TRBv1 has been tested and succesfully used in the new RPC
detector as a part of the upgrade of HADES detector system.

\section{TRB board version 2}
The enhanced version of TRB module (TRBv2) has been developed. The TRBv2 has
been designed as a multi-purpose readout module which could be used for all
types of detectors. 

\begin{figure}
  \begin{center}
    \includegraphics[width=0.6\textwidth]{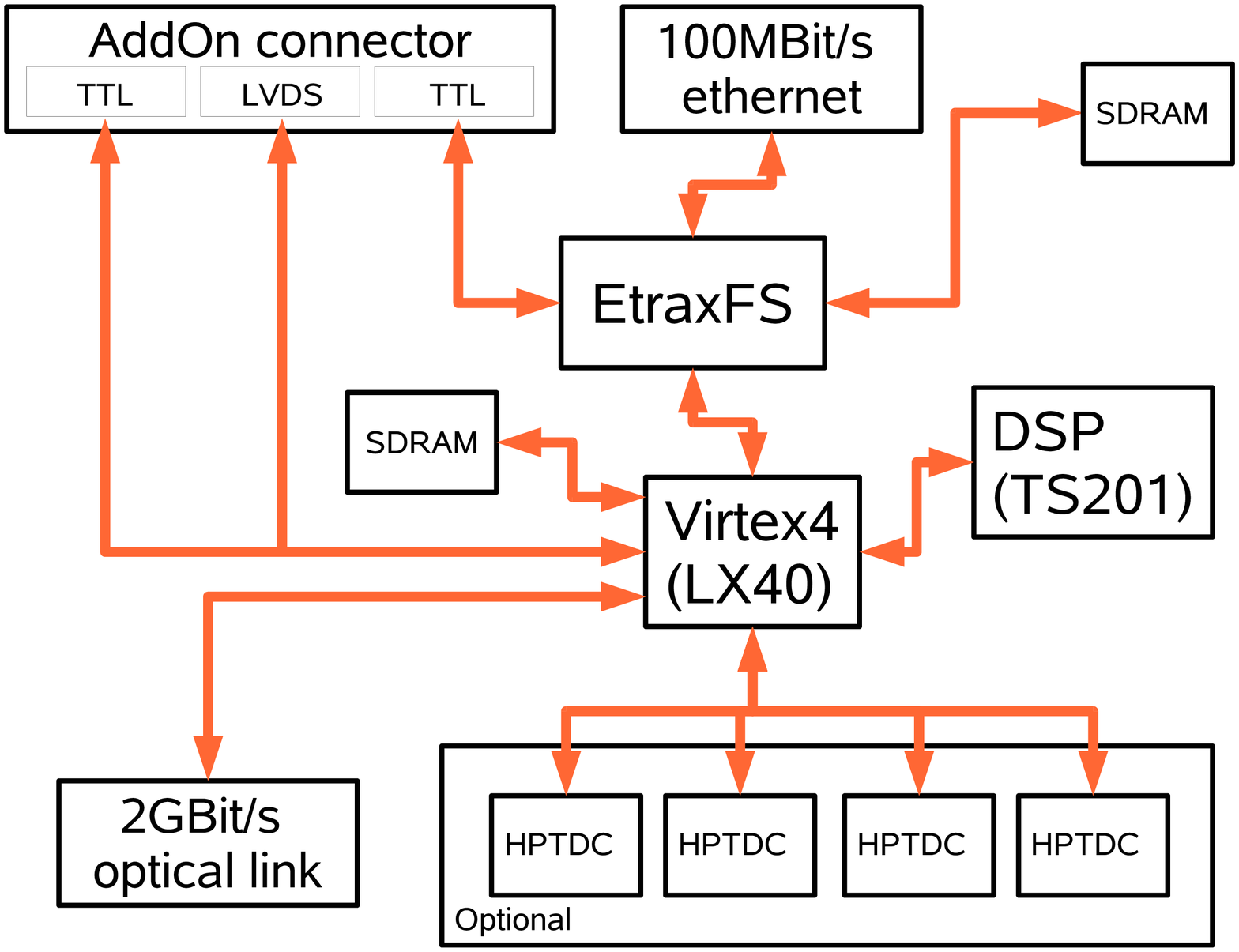}
\end{center}
\caption{TRBv2 block diagram. It consists of 4 HPTDC chips (optional),
  an Etrax-FS processor with 128 MB, an Ethernet interface, and a 2 Gbit/s
  optical link, a FPGA Virtex 4, and a TigerSharc DSP (Fig. taken from~\cite{tri}). }
\end{figure}

It uses an Etrax-FS processor running Linux for DAQ and
slow-control functionality.  It has three integrated I/O co-processors (each
200 MHz) which allow to use the bandwidth more efficentally without using too
much CPU computing resources, thus eliminating the bottelneck observed in the
TRBv1.  The TRBv2 has in addition an optical link 2 Gbit/s which can be used
for data transmission in the future applications (e.g. PANDA and CBM).  The
FPGA in the TRBv2 has been upgraded to a Xilinx Virtex 4 LX40 with 128 MB RAM.
Moreover, the new version provied an optional TigerSharc DSP which may
be used to implement on-line analysis algorithms.  To broaden the spectrum of
the applications, the TRBv2 is equipped with a high data-rate digital interface
connector (32 LVDS lines, 15 GBit/s) which allows to connect add-on boards
with the TRB.  The concept of add-on boards is explained in the next section.

\section{Add-on cards}

The TRBv2 has been designed as a detector-independent readout module.  All
detector-specific FEE, interfaces and connectors are placed on the add-on
cards which can be easily mounted on the TRB.  Additional computing resources
such as FPGA or DSP could be added to the add-on card.  The first add-on card
for the HADES-MDC (``Multiwire Drift Chambers'') readout has been
developed~\cite{attilio}. It has 10 connectors. Each connector has 50 pins and
two RS485 transcivers.  The operation of the MDC-add-on card is controlled by
the FPGA (Xilinx, XC4VLX-10FF1148) chip. The MDC-add-on board is connected back
to back to the TRBv2 via two connectors (SAMTEC QSE-040-01, 80 pins each) and
allows to read out each MDC-chamber by one single TRB board.  The concept of
add-on boards provides a possibility to mount specific-detector resources in a
flexible way, using the same base readout module for each detector.

\section{Conclusion}

The multi-purpose trigger and readout board with on-board DAQ functionality
(TRB) has been presented. The TRB has been designed for the HADES readout
system.  However, it can be used also in new CBM and PANDA detector systems
which will operate at FAIR facility at GSI.  The TRB version 1 has been
designed as a 128-channel Time to Digital Converter (TDC) based on the HPTDC
chip.  The module has been successfully implemented to readout the RPC
detector. As it has already the TDC chips on-board, it has been also used for
the prototype drift chamber readout tests for PANDA.  The TRB version 2 has an
2~GBit/s optical link (also very important for PANDA), add-on card connectors,
as well as an optional TigerSharc DSP.  Add-on boards for MDC, PreShower and
for other types of detectors are currently being tested.  It is planned to use
one more add-on board to serve for the readout the MAPS sensors for the CBM
experiment~\cite{schrader}.  Thus, the TRB module is proposed as a usefull
componant for new FAIR experiments.

\section*{Acknowledgments}

This work has been supported by the EU under the contracts CNI DIRAC-PHASE-1
(515876) and Hadron Physics (RII3-CT-2004-506078), the BMBF, the Helmholtz
Research School, the Polish MEEN (158/E-338/SPB/6.PR UE/DIE 455/2004-2007) and
GSI.

\end{document}